\documentclass[journal]{IEEEtran}
\usepackage{amsmath}
\usepackage[mathscr]{euscript}
\usepackage[pdftex]{graphicx}
\usepackage[nocompress]{cite}

\begin{document}

\title{Achieving Super-Resolution \\with Redundant Sensing}

\author{	Diu~Khue~Luu$^*$,
			Anh~Tuan~Nguyen$^*$,~\IEEEmembership{Student Member,~IEEE,}
			and~Zhi~Yang$^1$,~\IEEEmembership{Member,~IEEE,}
\thanks{$^*$Co-first authors. $^1$E-mail: yang5029@umn.edu.}
\thanks{Department of Biomedical Engineering, University of Minnesota, Minneapolis, MN 55455, USA.}}


\maketitle

\begin{abstract}

Analog-to-digital (quantization) and digital-to-analog (de-quantization) conversion are fundamental operations of many information processing systems. In practice, the precision of these operations is always bounded, first by the random mismatch error (ME) occurred during system implementation, and subsequently by the intrinsic quantization error (QE) determined by the system architecture itself. In this manuscript, we present a new mathematical interpretation of the previously proposed redundant sensing (RS) architecture that not only suppresses ME but also allows achieving an effective resolution exceeding the system's intrinsic resolution, i.e. super-resolution (SR). SR is enabled by an endogenous property of redundant structures regarded as ``code diffusion'' where the references' value spreads into the neighbor sample space as a result of ME. The proposed concept opens the possibility for a wide range of applications in low-power fully-integrated sensors and devices where the cost-accuracy trade-off is inevitable.

\end{abstract}


\IEEEpeerreviewmaketitle

\section{Introduction}

\IEEEPARstart{T}{he} process of \textit{quantization} i.e. analog-to-digital conversion (ADC) and the reverse operation \textit{de-quantization} i.e. digital-to-analog conversion (DAC) are the basis of all modern sensory data acquisition systems. They allow ``digital'' artificial systems to sense and interact with the ``analog'' physical world. Quantization is essentially a lossy data compression process where information from a higher-resolution space is represented in lower-resolution counterpart. In practical implementations, the precision of this process is always bounded by the system resource constraints such as size, power, bandwidth, and memory, etc. For example, in many integrated circuits ADC and DAC designs, an addition 1-bit of resolution or 2x precision often require a 4x increase of chip area and power consumption \cite{2008_Murmann, 2015_Fredenburg, 2015_Murmann}. While ultra-high resolution ADCs/DACs up to 32-bit is possible, the large size and power consumption limit the use of these devices in many practical applications. Similarly, higher resolution image sensor requires more pixel count and buffer memory thus also results in larger device and power consumption. While it is possible to improve the pixel density, the smaller pixel size is associated with increased noise which limits the sensor's dynamic range \cite{2005_El, 2014_Fossum}.

Super-resolution (SR) are techniques that aim at achieving an effective resolution exceeding the precision that the system's resource constraints commonly permit. They have wide applications various in fields of engineering and science concerning imaging and instrumentation where higher resolution data acquisition is always desired \cite{2003_Park, 2011_Tian}. Previous SR techniques focus on recovering fine details of the object of interest by integrating the information obtained from coarse observations. These techniques could be generally divided into two primary classes: modeling-based and oversampling-based, which are also known as single-frame and multi-frame in image processing. Modeling-based (single-frame) techniques such as \cite{2002_Freeman, 2003_Tipping, 2004_Chang, 2009_Glasner, 2010_Yang} focus on modeling the input sources from available data points and reconstructing the missing information by means of approximation. On the other hand, oversampling-based (multi-frame) techniques such as \cite{2004_Farsiu, 2009_Lu, 2010_Li, 2015_Huang} acquire and combine multiple samples of the input obtained at various spatial or temporal instants to extract the sub- least-significant-change information.

In this manuscript, we present a new approach based on the redundant sensing (RS) \cite{2015_Nguyen, 2016_Nguyen_NIPS, 2018_Nguyen} that requires neither modeling nor oversampling of the input signals. A RS structure is essentially a redundant system of information representation where each outcome in the sample space can be generated by multiple distinct system configurations. In practice, these configurations are always affected by random mismatch error (ME), which conventionally, is considered as a ''problem`` causing conversion error and degrading the system's overall precision. Yet here we show that ME allows actual values of the system's redundant configurations to ``diffuse'' into the neighbor sample space such that with a sufficient level, a RS structure has the information capability to quantize the data at an effective resolution beyond the conventional resource constraints. 

Our SR technique is fundamentally different from previous approaches because it does not involve reconstructing the missing information. Instead, the mechanism aims at direct sampling of higher-precision data points provided the quantizer's endogenous structure is correctly optimized. Indeed, the fine-detailed information content of the input signal is never lost during quantization. This is achieved not only because of the RS architecture itself but also by elegantly manipulating ME - an undesirable precision-limiting factor in conventional designs. 

In the following sections, the background and mechanisms which facilitate SR in a RS architecture are explained. We use Monte Carlo simulations to demonstrate an extra 8-9 bits resolution or 256-512x precision can be accomplished on top of a 10-bit quantizer at 95\% sample space. Lastly, we discuss potential applications and practical considerations of the proposed technique in fully-integrated miniaturized biomedical devices where the structure's complexity can be mitigated by approximation or conveniently circumvented.

\section{Super-Resolution}
\label{Sec_SuperRes}

\subsection{Quantization \& Mismatch Error}
\label{Sec_Quantization}

Quantization\footnote{De-quantization is defined similarly and share the same characteristics.} is a process of mapping a continuous set (analog) to a finite set of discrete values (digital). Without loss of generality, we can assume a $N_0$-bit quantizer divides the continuous interval [0, 1) into $2^{N_0}$ partitions defined by a set of \textit{references} $\theta_0 \leq \theta_1 \leq ... \leq \theta_{2^{N_0}}$ where each partition is mapped onto a digital code $d$ ranging from 0 to $2^{N_0}-1$:
\begin{equation}
x_A \in [\theta_d, \theta_{d+1}) \rightarrow x_D = d, \quad \forall d=0, 1, ..., 2^{N_0}-1
\end{equation}
where $x_A$ is the analog input and $x_D$ is the digital output. The quantizer's \textit{effective resolution} can be quantified by the \textit{Shannon entropy} $H_{N_0}$ as follows:
\begin{equation}
\label{Eq_ShannonEntropy}
\begin{split}
    M_{N_0} &= \sum_{d=0}^{2^{N_0}-1} \int_{\theta_d}^{\theta_{d+1}} { \left( x_A - \frac{d+0.5}{2^{N_0}} \right)^2 d x_A } \\
    H_{N_0} &= -\log_2 \sqrt{12 \cdot M_{N_0} }
\end{split}
\end{equation}
where $M$ is the normalized total mean-square-error integrated over each digital code. It can be shown that $H_{N_0} \leq N_0$ for all values of reference $\theta_d$. Equality occurs only when $2^{N_0}$ references are equally spaced, i.e. $\forall i, j: \theta_{i+1}-\theta_{i} = \theta_{j+1}-\theta_{j}$. This fundamental maximum value of entropy is referred as the \textit{Shannon limit}, where the device's resolution is bounded only by its intrinsic quantization error (QE).

In practice, the quantizer's precision is also affected by the randomly occurred ME resulting in the undesirable deviation of the references and degradation of entropy. For example, integrated circuits ADC or DAC chips such as \cite{2015_Nguyen} generate their references by arrays of identical elementary components regarded simply as \textit{unit cells}. A $N_0$-bit device generally has $2^{N_0}-1$ unit cells which could be miniature capacitors, resistors or transistors. The random mismatch of individual unit cells due to variations of the fabrication process and other non-ideal factors is one of the primary sources of ME that could significantly deteriorate the device's precision \cite{1989_Pelgrom, 2005_Croon, 2012_Fredenburg}.

To effectively control the unit cells, they are always grouped into bundles regarded simply as \textit{components}. Grouping significantly reduces the number of control signals required. For example, with the conventional binary-weighted method, $2^{N_0}-1$ unit cells are arranged into $N_0$ components with the nominal weight of $\{ 2^0, 2^1,..., 2^{N_0-1} \}$. Such system is \textit{orthogonal} because with $N_0$ binary control signals, i.e. 0/1 bits, $2^{N_0}$ references corresponding to each digital code in $[0, 2^{N_0}-1]$ can be uniquely created by selecting and assembling the components according to the binary numeral system. 

\subsection{Redundant Sensing}
\begin{figure}[t]
\begin{center}
\includegraphics[width=0.48\textwidth]{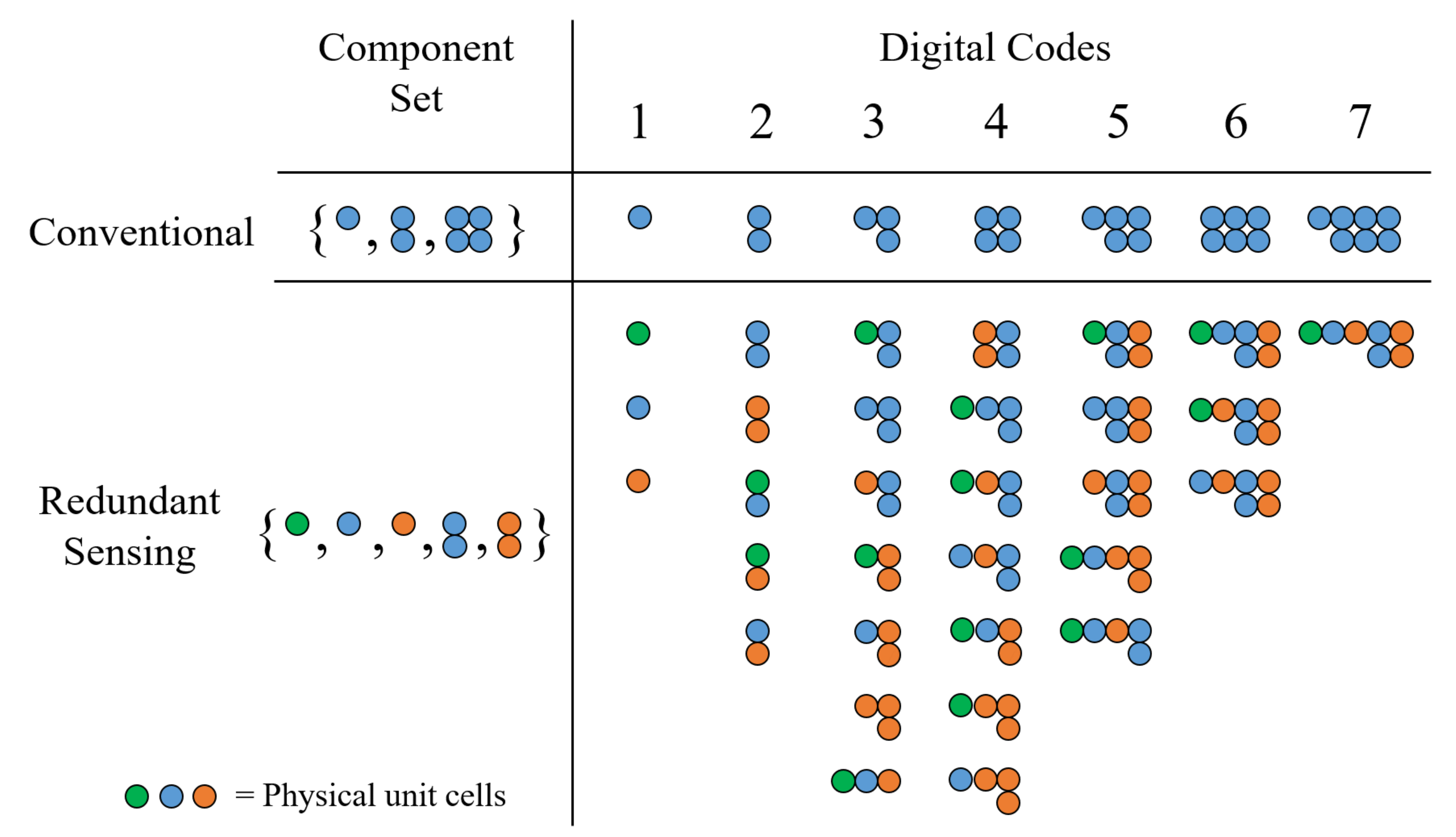}
\end{center}
\vspace{-0.1in}
\caption{Illustration of a simple 3-bit redundant sensing (RS) structure where representational redundancy (RPR) and entangled redundancy (ETR) can be achieved by utilizing a non-orthogonal grouping method without the need for replication. While using the same amount of physical resource (i.e. 7 unit-cells), in the RS structure, each digital code can be created by multiple distinct assemblies of components, each expresses a different, partially correlated distribution with respect to random ME.}
\label{Fig_RedunSense}
\end{figure}

RS is a design framework that aims at engineering redundancy for enhancing the system's performance regarding accuracy and precision, instead of reliability and fault-tolerance like conventional designs\cite{2016_Nguyen_NIPS}. A practical RS implementation must satisfy two criteria, namely \textit{representational redundancy} (RPR) and \textit{entangled redundancy} (ETR) \cite{2018_Nguyen}. 

RPR refers to a \textit{non-orthogonal} scheme of information representation where every outcome in the sample space is encoded by numerous distinct system configurations. Each configuration responses differently to ME such that in any given instance, there almost always exists one or more configurations that have smaller errors than the conventional representation. 

ETR refers to the implementation of the RS structure such that the statistical distribution of different system configurations is partially correlated (i.e. entangled) allowing a large degree of redundancy without incurring excessive resource overhead. ETR should be differentiated from conventional replication-based method to realize redundancy where the degree of redundancy is linearly proportional to the resource utilization. 

Fig. \ref{Fig_RedunSense} illustrates a simple example where a 3-bit RS structure with both RPR and ETR properties can be accomplished by utilizing a non-orthogonal grouping method without the need for replication. While using the same amount of physical resource (i.e. 7 unit cells), in the RS structure, each digital code can be created by multiple distinct assemblies of components, each expresses a different, partially correlated distribution with respect to random ME. This redundant system of information representation has been shown to suppress ME by allowing searching for the optimal component assembly with the least error with respect to each and every digital code \cite{2016_Nguyen_NIPS}. In this work, we will show that such redundant mechanism can be elegantly exploited to realize an effective resolution beyond the conventional limit of $N_0$ bounded by QE. 

\subsection{Code Diffusion}
\label{Sec_CodeDiff}
\begin{figure*}[t]
\begin{center}
\includegraphics[width=\textwidth]{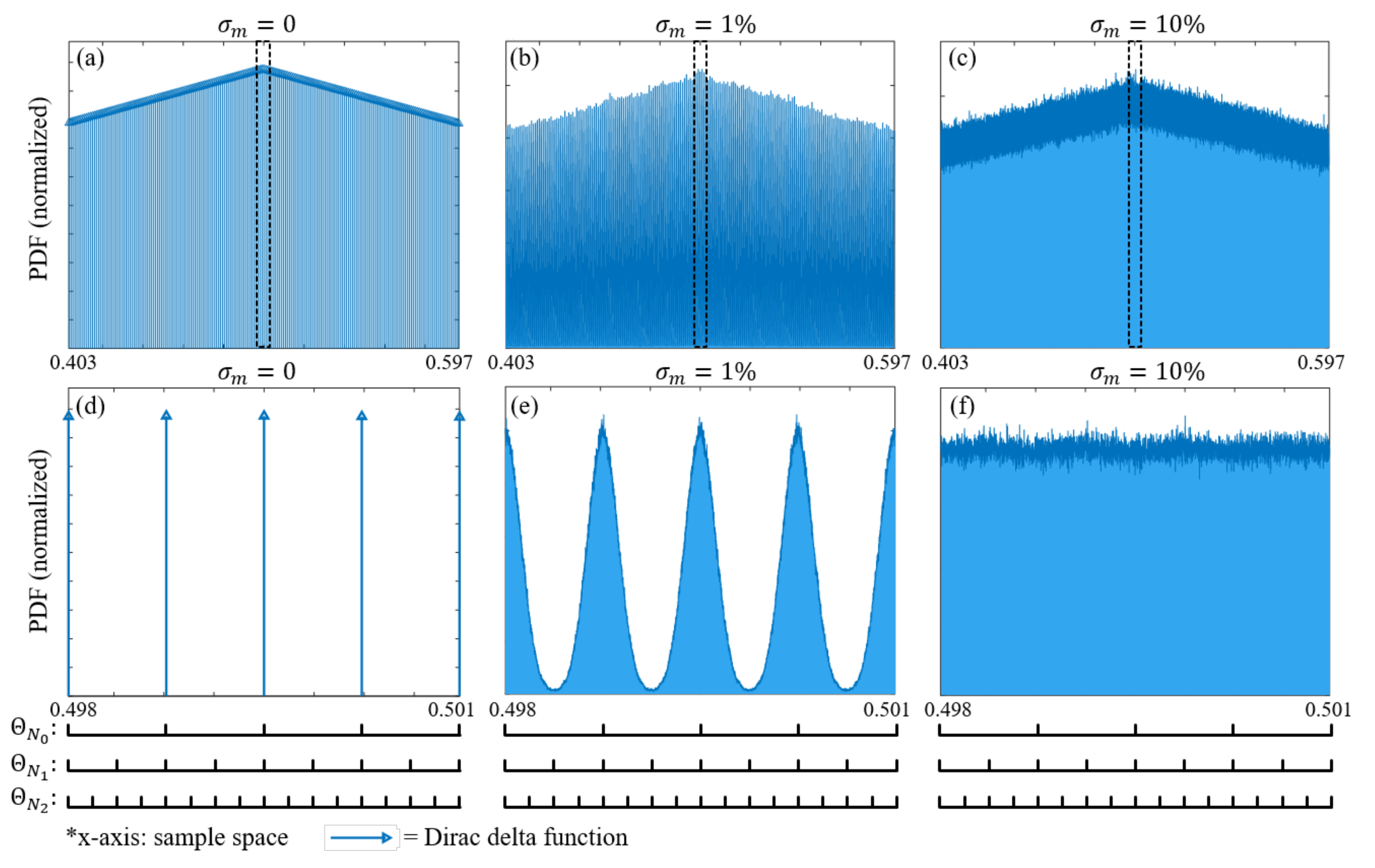}
\end{center}
\vspace{-0.1in}
\caption{(a-c) Estimated probability density function (PDF) of all the references $\{ \theta_0, \theta_1, ... \}$ that can be generated by an example RS structure at various mismatch ratio $\sigma_m$ (0-10\%) and (d-f) their respective zoom-in views which show the PDF's segments centered at each element of $\Theta_{N_0}$. With sufficient level of mismatch ratio, the references ``diffuse'' evenly across the sample space which allows approximating elements of a higher-resolution set $\Theta_{N_1}, \Theta_{N_2}, ...$ and facilitating SR.}
\label{Fig_CodeDiff}
\end{figure*}

Fig. \ref{Fig_CodeDiff} shows the estimated probability density function (PDF) of all the references $\{ \theta_0, \theta_1, ... \}$ that can be generated by an example RS structure at various \textit{mismatch ratio}. The mismatch ratio $\sigma_m$ is defined as the standard deviation of each unit cell which is assumed to have a Gaussian distribution with unity mean. Monte Carlo simulations ($n=1000$) are performed on a redundant structure similar to one implemented in \cite{2015_Nguyen} with $N_0=10$.

Clearly, in the absence of ME or $\sigma_m = 0$ (Fig. \ref{Fig_CodeDiff}a, d), regardless how the unit cells are grouped and assembled, an array of $2^{N_0}-1$ identical units can only generate a finite number of references belonging to the following discrete set of values:
\begin{equation}
\Theta_{N_0} = \{ \frac{0}{2^{N_0}}, \frac{1}{2^{N_0}}, ..., \frac{2^{N_0}-1}{2^{N_0}}, \frac{2^{N_0}}{2^{N_0}} \}
\end{equation}
$\Theta_{N_0}$ is regarded as the \textit{intrinsic reference set} (IRS) corresponding to an effective resolution $N_0$. The elements of $\Theta_{N_0}$ are marked by Dirac delta functions in Fig. \ref{Fig_CodeDiff}a, d.

As $\sigma_m$ assumes non-zero values (Fig. \ref{Fig_CodeDiff}b, e), the PDF's segment centered at each element of $\Theta_{N_0}$ is widen as the actual values generated by different component assemblies begin ``diffusing'' into the neighbor sample space. This property is unique to a RS structure because (i) there are numerous different component assemblies that can generate references with the same nominal values, i.e. RPR, and (ii) the distribution of these assemblies are partially independent with respect to random ME, i.e. ETR. Subsequently, the spreading of the PDF occurs at every trial of ME, not merely the result of the Monte Carlo sampling.

In an ordinary quantizer, code diffusion is undesirable because it makes the references deviates from $\Theta_{N_0}$, thus results in the degradation of the Shannon entropy as shown in equation \ref{Eq_ShannonEntropy}. In fact, our previous system in \cite{2016_Nguyen_NIPS} was designed to reverse the diffusing process by searching for the assemblies that are closest to each element of $\Theta_{N_0}$.

However, from another perspective, code diffusion implies that the same system could generate references within the sample space' regions that are belonged to the IRS of a higher resolution $N_k = N_0+k$: 
\begin{equation}
\Theta_{N_k} = \{ \frac{0}{2^{N_k}}, \frac{1}{2^{N_k}}, ..., \frac{2^{N_k}-1}{2^{N_k}}, \frac{2^{N_k}}{2^{N_k}} \}
\end{equation}
where $\Theta_{N_0} \subset \Theta_{N_1} \subset ... \subset \Theta_{N_k}$ ($k = 1, 2,...$) as marked in the x-axis of Fig. \ref{Fig_CodeDiff}d, e, f. With sufficient level of mismatch ratio (Fig. \ref{Fig_CodeDiff}c, f), the reference's PDF covers almost all the sample space with relatively even chances. Subsequently, there is an adequate possibility a set of assemblies closely approximating $\Theta_{N_k}$ can be found that would allow sampling at an effective resolution $N_k$ beyond the system intrinsic resolution $N_0$. It is also interesting to point out that ME, which is conventionally regarded as an undesirable non-ideal factor, is the crucial element that enable SR. Maximal effectiveness of SR is obtained only when the mismatch ratio reaches a certain level $\sim$10\% which is considered excessive large in many ordinary applications.

Such mechanism is only possible because the number of distinct references that can be generated by a RS structure is significantly larger than the cardinality of both $\Theta_{N_0}$ and $\Theta_{N_k}$ due to redundancy. In an orthogonal structure such as the binary system, the number of distinct references is strictly $2^{N_0} = |\Theta_{N_0}|$, which is smaller than $|\Theta_{N_k}|$ for all $k$. Furthermore, not only the number of different component assemblies but also the mutual correlation between them play an important role. Ideally, we want the assemblies to spread evenly across all the sample space to have the maximum chance of approximating $\Theta_{N_k}$. This characteristic is determined by the device's internal architecture, i.e. how the components are designed.

\subsection{Grouping Method}
\label{Sec_GroupingMethod}

The \textit{grouping method} (GM) is the way unit cells are arranged into components. Almost all conventional designs can be categorized as binary-weighted (BW) structures where the quantization partitions are uniquely encoded according to the binary numeral system. In contrast, the proposed RS architecture employs a different strategy to realize redundancy with both RPR and ETR properties. There is no limitation to how the unit cells are grouped. While GM does not alter the number of unit cells, thus has little effect on the resource constraints, it determines the system's endogenous architecture and greatly affects the references' number and distribution. The design of GM differentiates one redundant structure from another. 

Let assume a given GM assembles $2^{N_0}-1$ unit cells into $n$ components with the nominal weight $\bar{C} = \{ \bar{c}_1, \bar{c}_2, ..., \bar{c}_n \}$ and the actual weight $C = \{ c_1, c_2, ..., c_n \}$ with respect to random ME. Each subset of $C$, encoded by the binary string $d = \overline{d_1d_2...d_n}$ ($d_i \in \{0, 1\}$), generates a normalized reference $\theta_d$ as follows:
\begin{equation}
\theta_d = \Sigma_{i=1}^{n} d_ic_i / (1+\Sigma_{i=1}^{n} c_i)
\end{equation}
Let $\Phi$ is the set of all references that can be generated by system. To achieve an effective resolution $N_k$ is essentially to search for a subset $\hat{\Theta}_{N_k} \subset \Phi$ that closely approximates $\Theta_{N_k}$. Clearly, SR can only be accomplished in a redundant structure as $|\Phi| > |\Theta_{N_k}|$ or $n> N_k$.

\begin{figure}[t]
\begin{center}
\includegraphics[width=0.45\textwidth]{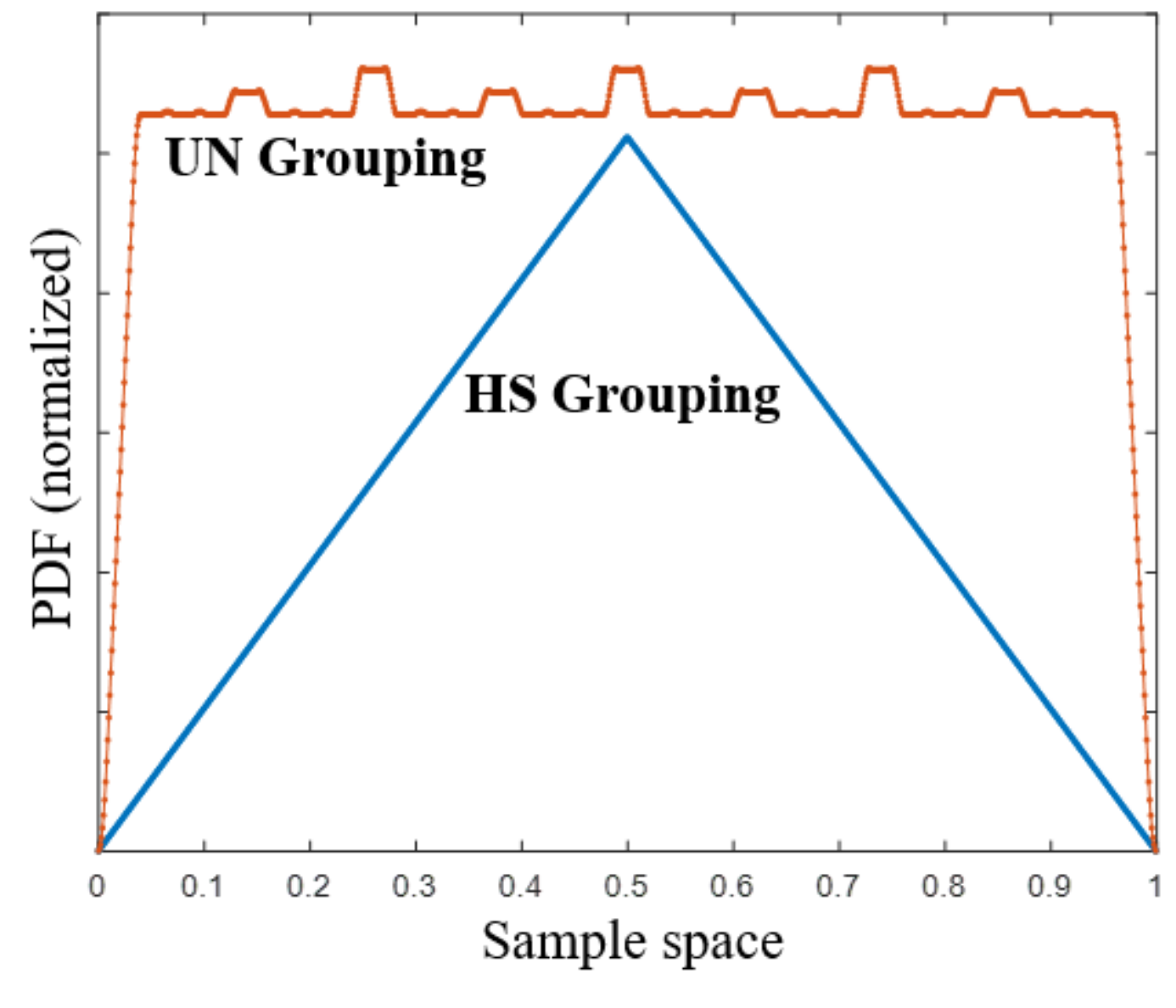}	
\end{center}
\vspace{-0.2in}
\caption{A comparison of the reference distribution between the HS and UN grouping method. The UN method yields more uniform distribution across different regions of the sample space, especially the two ends, which would translate to better SR potential.}
\label{Fig_GroupMeth}
\end{figure}

The previously proposed RS architecture employed a class of GM that was inspired by the binocular structure of the human visual system. They yield the nominal weight $\bar{C}_{\text{RS}} = \bar{C}_{\text{RS}, 0} \cup \bar{C}_{\text{RS},1}$ according to the following formula with parameters $(s, N_0')$ satisfied $1 \leq N_0' < N_0, 1 \leq s \leq N_0 - N_0'$:
\begin{equation}
\begin{split}
\bar{C}_{\text{RS},1} &= \{\bar{c}_{1, i} | \bar{c}_{1, i} = 2^{N_0-N_1+i-s} \}\\
\bar{C}_{\text{RS},0} &= \{\bar{c}_{0, j} | \bar{c}_{0, j} =
	\begin{cases}
    2^j, \text{ if } j<N_0-N_0'\\
    2^j-\bar{c}_{1, j-N_0+N_0'}, \text{ otherwise}
  	\end{cases} \}
\end{split}
\end{equation}
where $i \in [0, N_1-1]$, $j \in [0, N_0-1]$. The special case of $\bar{C}_{\text{RS}}$ where $N_0' = N_0-1$ and $s=1$ called the ``half-split'' (HS) array has been demonstrated in \cite{2015_Nguyen}. It has the following nominal weight $\bar{C}_{\text{HS}} = \bar{C}_{\text{HS},0} \cup \bar{C}_{\text{HS},1}$ where:
\begin{equation}
\begin{split}
\bar{C}_{\text{HS},0} &= \{ 2^0, 2^1, ..., 2^{N_0-2} \} \cup \{ 2^0 \}\\
\bar{C}_{\text{HS},1} &= \{ 2^0, 2^1, ..., 2^{N_0-2} \}
\end{split}
\end{equation}
Among the RS structures, the HS design has the largest number of components thus the greatest degree of redundancy while contains a reasonable number of components of $2N_0-1$. Also, the simplicity of the design allows it to be implemented in hardware with minimal complexity as presented in \cite{2015_Nguyen}. The distribution of $\Phi_{HS}$ is shown in Fig. \ref{Fig_GroupMeth}. While the HS method has a high level of redundancy, their distributions are not necessarily optimal for achieving SR. The references mostly concentrate into the middle region of the sample space leaving the two ends inadequately covered and vulnerable to errors. 
\begin{figure*}[t]
\begin{center}
\includegraphics[width=\textwidth]{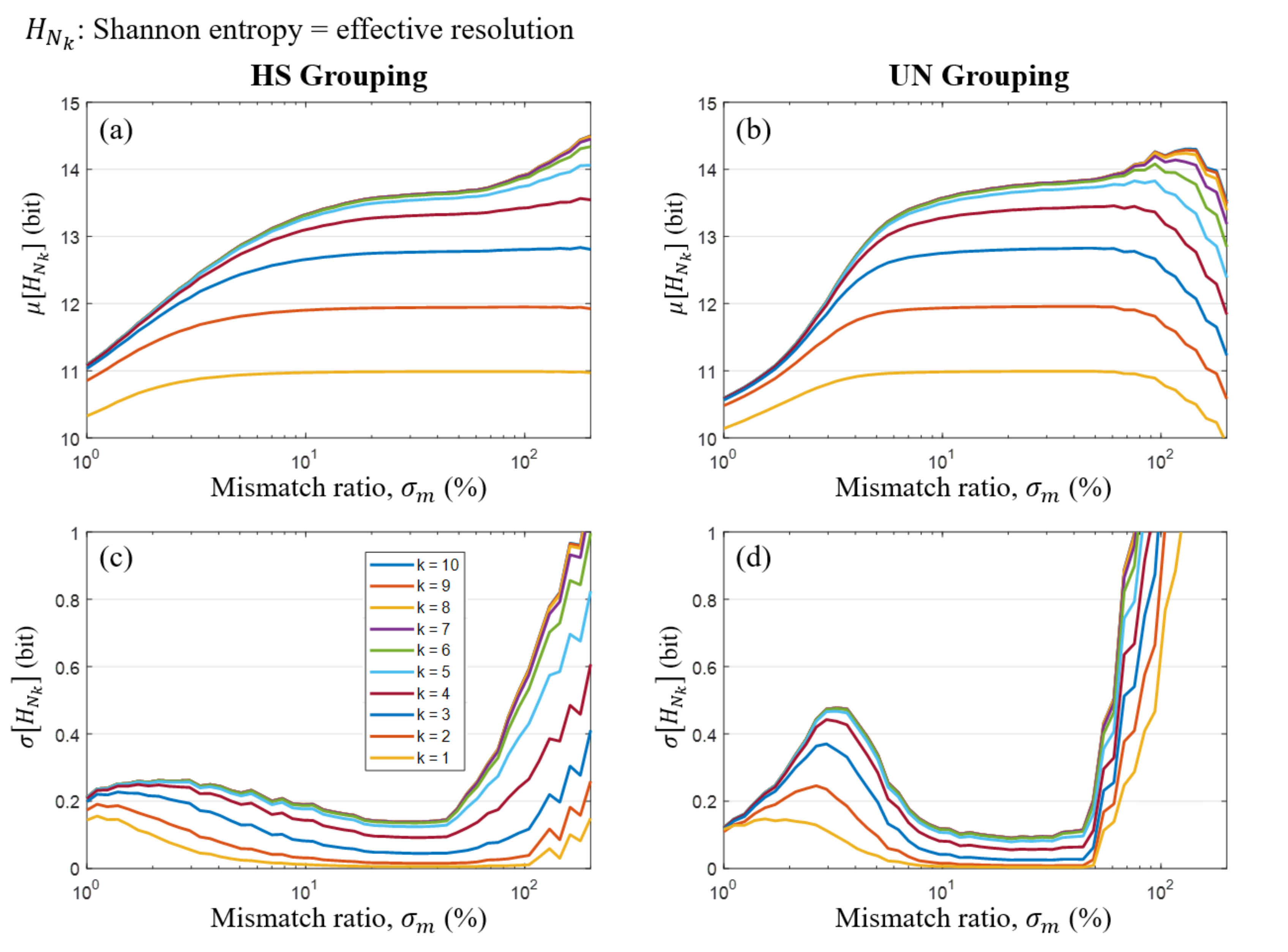}	
\end{center}
\vspace{-0.2in}
\caption{Achievable SR in HS and UN redundant structures: (a, b) mean ($\mu[H_{N_k}]$) and (c, d) standard deviation ($\sigma[H_{N_k}]$) of the entropy of a $N_0=10$ bits device at various targeted resolution $N_k = N_0+k$ and mismatch ratio $\sigma_m$. With sufficient mismatch ratio, 3-4 bits increase of effective resolution or 8x-16x enhancement of precision.}
\label{Fig_SupRes}
\end{figure*}

In this manuscript, we propose an enhanced GM that is specifically designed to support SR. It has a more uniform distribution of references to maximize the coverage of the sample space. The ``UNiform'' (UN) method yields the following nominal weight $\bar{C}_{\text{UN}} = \bar{C}_{\text{UN}, 0} \cup \bar{C}_{\text{UN},1} \cup ... \cup \bar{C}_{\text{UN}, \lfloor log_2{N_0} \rfloor}$ where:
\begin{equation}
\begin{split}
	\bar{C}_{\text{UN},i} &= \{ \bar{c}_{i,j} | \bar{c}_{i,j} = 2^j \} \\
	\bar{C}_{\text{UN}, 0} &= \{ \bar{c}_{0,l} | \bar{c}_{0,l} = 
	\begin{cases}
	    2^l, \text{ if } l<N_0-N_1 \\
    	2^l - \sum\limits_{m=1}^{\lfloor log_2{N_0} \rfloor} 2^{l-N_0+N_m}, \text{ otherwise}
  	\end{cases} \}
\end{split}
\end{equation}
where $N_i = \lceil N_{i-1}/2 \rceil$ $\forall i \in [1, \lfloor log_2 N_0 \rfloor ]$, $j \in [0, N_i-1]$, $ l \in [0, N_0-1]$. The intuition behind the UN design is to divide the components of a binary-weighted array into numerous sub-arrays with different resolutions $N_1, N_2, ...$ that reduce in log scale. This maximizes the distribution of small and large components over the digital codes while retaining the total number of components at a reasonable value of $~2N_0$ similar to the HS structure. All the remaining components form the base array $\bar{C}_{\text{UN}, 0}$.

As a comparison, with $N_0 = 10$, the BW, HS and UN methods yield the following nominal component set:
\begin{equation}
\begin{split}
	&\bar{C}_{\text{BW}} = \{ 1, 2, 4, 8, 16, 32, 64, 128, 256, 512 \} \quad (10 \text{ elements}) \\
	&\bar{C}_{\text{HS}} = \{ 1, 1, 1, 2, 2, 4, 4, 8, 8, 16, 16, 32, 32,
		\\& \qquad \qquad \qquad \quad 64, 64, 128, 128, 256, 256 \} \quad (19 \text{ elements})\\
	&\bar{C}_{\text{UN}} = \{ 1, 1, 1, 1, 2, 2, 2, 2, 4, 4, 4, 8, 8, 16, 
		\\& \qquad \qquad \qquad \qquad 16, 31, 62, 123, 245, 490 \} \quad (20 \text{ elements})\\
\end{split}
\end{equation}
As shown in Fig. \ref{Fig_GroupMeth}, the UN method gives significantly ``flatter'' distribution of references which would translate to more even code diffusion over different regions of the sample space. In the following sections, we will show this property helps suppress errors near the two ends of the sample space and results in more SR potential in general.
 	
\subsection{Beyond The Shannon Limit}
\label{Sec_ShannonLim}

SR in the context of this work should be understood as a resource-constraint problem. The precision of a sensor consists of $2^{N_0}-1$ unit cells was previously thought to be bound by the Shannon limit of $N_0$ determined by QE. By arranging the unit cells in a specific manner to realize a redundant structure and exploiting the statistical property of random ME, we aim to achieve an effective resolution beyond this conventional ``limit''.
\begin{figure*}[t!]
\begin{center}
\includegraphics[width=\textwidth]{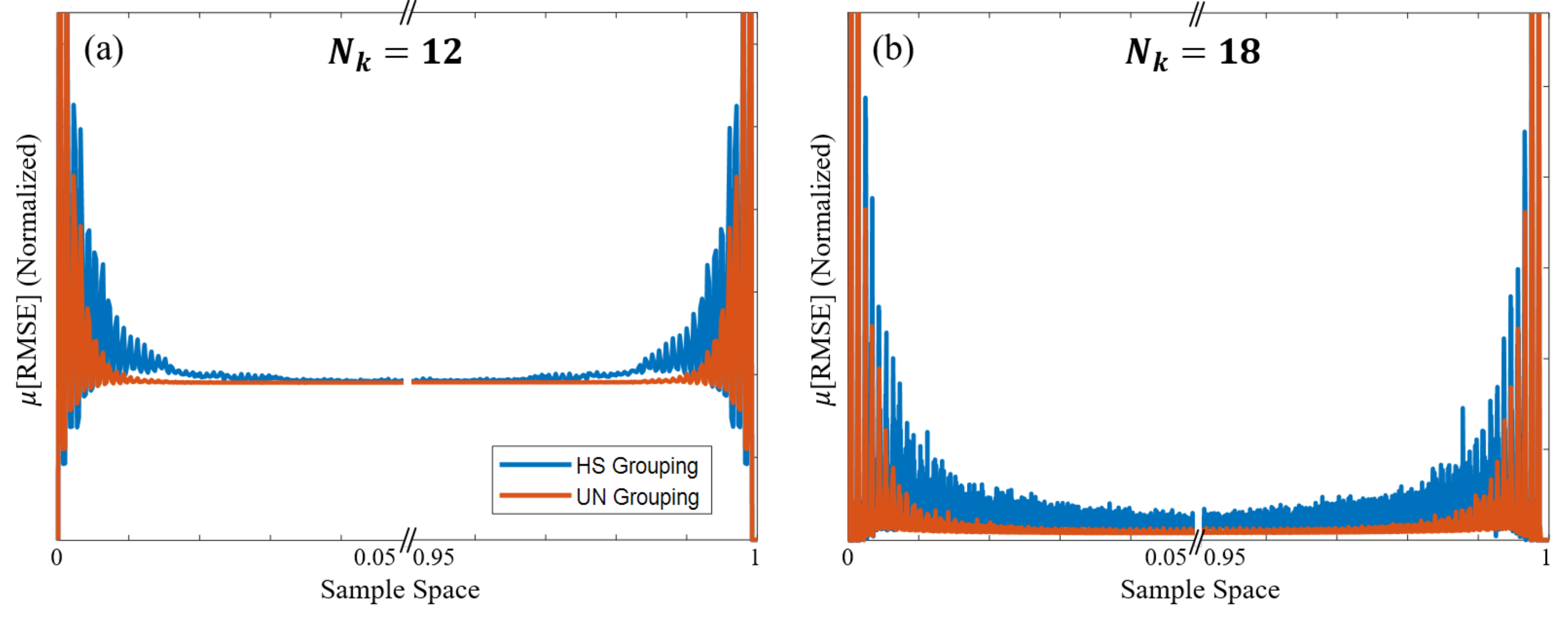}	
\end{center}
\vspace{-0.2in}
\caption{Root mean square error (RMSE) computed over the sample space ($N_0 = 10$, $\sigma_m = 10\%$). Note that the x-axis only shows the first and last 5\% of the sample space. At a high-resolution, errors mostly occur at the two ends where the level of redundancy is lower. This can result in significant degradation of the overall entropy. The UN method is designed to have flatter code distribution which helps shape the error to the extreme end.}
\label{Fig_RMSE}
\end{figure*}

The Shannon limit exists because the ordinary expression of entropy as shown in Equation \ref{Eq_ShannonEntropy} is computed against a reference set of only $2^{N_0}+1$ values $\{ \theta_0, ..., \theta_{N_0} \}$ which is the maximum number of distinct references a conventional binary-weighted array can generate. This limitation does not apply to a redundant architecture. A HS or UN structure has a reference set $\Phi_{\text{HS}}/\Phi_{\text{UN}}$ with as much as $\sim 2^{2N_0}$ distinct elements. The key to achieve SR is to find a subset $\hat{\Theta}_{N_k}$ from $\Phi_{\text{HS}}/\Phi_{\text{UN}}$ such that $\hat{\Theta}_{N_k}$ closely approximates the IRS $\Theta_{N_k}$ at the resolution $N_k$. This can only be accomplished when there is random ME that allows the elements of $\Phi_{\text{HS}}/\Phi_{\text{UN}}$ to diffuse across the sample space. Hence, the concept of SR does not contradict with the conventional Shannon limit, but a new interpretation of the Shannon theory beyond its ordinary understanding that only applies in a practical redundant architecture.

The Shannon entropy in Equation \ref{Eq_ShannonEntropy} can be conveniently modified to represent the effective resolution at a targeted resolution $N_k$ by replacing $N_0 \leftarrow N_k$ and extend the scope to $\theta_d$ to include all the values in $\hat{\Theta}_{N_k}$. Fig. \ref{Fig_SupRes} show the mean and standard deviation (STD) of the estimated entropy of a $N_0=10$ bit device using Monte Carlo simulations ($n=1000$) at various targeted resolution and mismatch ratio. The optimal set $\hat{\Theta}_{N_k}$ is found using exhaustive search.

As our analysis of code diffusion suggested, the best performance of SR is obtained with the mismatch ratio above $\sim 10\%$. Both HS and UN grouping method offers 3-4 bits increase of effective resolution or 8x-16x enhancement of precision. The entropy's STD is less than $0.2$-bit within 10-50\% mismatch ratio where the UN method has a marginally better outcome. These results suggest that the solution for SR is consistent which in practical applications, will translate to the good yield of the device under random error. 

Furthermore, the consistency of the mechanism implies that ME may not need to be truly ``random''. In certain application, 10\% random deviation may seem unrealistic. Instead, the deviation can be intentionally added to the structure during the design process. Even if these artificial pseudo-random deviations could carry a certain level of error, the consistency of SR mechanism guarantees that a solution can always be found.

\subsection{Reduced-Range Sampling}
\label{Sec_ReducedRange}
\begin{figure*}[t!]
\begin{center}
\includegraphics[width=\textwidth]{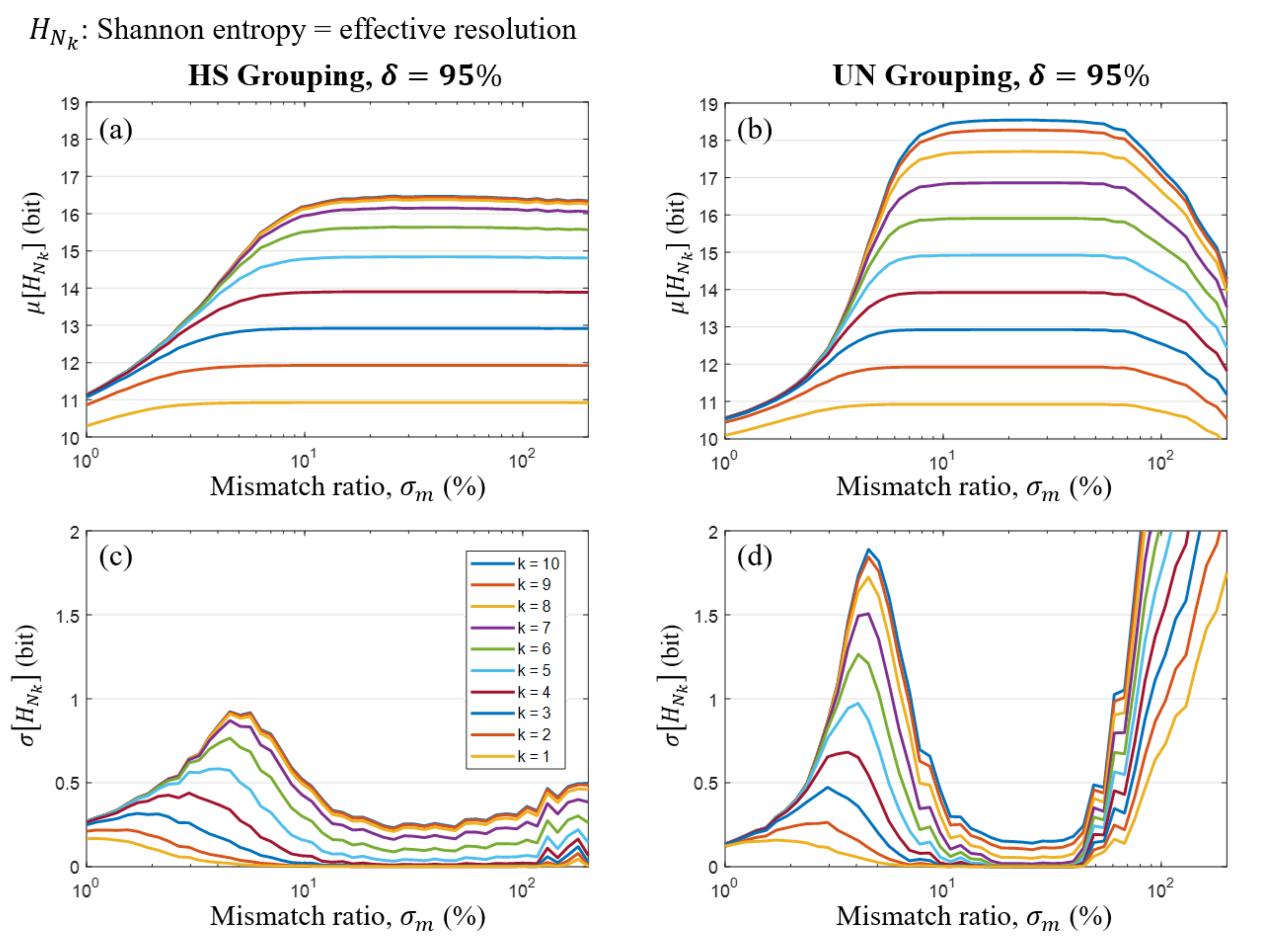}	
\end{center}
\vspace{-0.2in}
\caption{Achievable SR of the same HS and UN redundant structures in Fig. \ref{Fig_SupRes} ($N_0=10$, $N_k = N_0+k$) but at $\delta = 95\%$ sample space. By sacrificing 5\% of the sample space - a reasonable engineering trade-off, an increase of 8-9 bits effective resolution or 256x-512x enhancement of precision is feasible with the UN structure.}
\label{Fig_ReducedRange}
\end{figure*}

Fig. \ref{Fig_RMSE} shows the distribution of the root mean square error (RMSE) over the sample space or the value of $\sqrt{M_{N_k}(d)}$ at each digital code $d$ in Equation \ref{Eq_ShannonEntropy} before the summation. At a high-resolution, errors mostly occur at the two ends of the sample space. These are regions that have a lower level of redundancy as implied by the code distribution presented in Fig. \ref{Fig_GroupMeth}. The UN technique is designed to have better spreading of the codes compared to the HS design, thus help mitigate parts of the error by shaping it to the extreme ends. However, because of the nature of the grouping, it is mathematically not possible to cover all the sample space equally. 

Nevertheless, we argue that many applications actually do not utilize the entire sample space equally due to numerous practical reasons. The majority of sensors are calibrated such that the signals that need to be captured fall within the middle of the sample space. This is because most signals do not distribute uniformly across the sample, ``centering'' the data minimize the chance of the signal going beyond the sampling range causing distortion and loss of information. Suppose we can simply ignore the two extreme ends, the proposed technique allows realizing a continuous sampling range centered at the middle of the sample space where the overall effective resolution can be significantly enhanced.

Let $\delta \in [0, 1]$ is the length of a continuous region centered at the middle of the sample space where data are captured. This effectively reduces the full-range and dynamic range of the device which results in a lower Shannon limit:
\begin{equation}
 \max(H_{N_k, \delta}) = \log_2(\delta 2^{N_k})=  N_k+ \log_2 \delta
\end{equation}
The normalized total mean square error and entropy are now only integrated over a smaller range of digital codes:
\begin{equation}
\begin{split}
    M_{N_k, \delta} &= \sum_{d=\lfloor \frac{1-\delta}{2} \cdot 2^{N_k} \rfloor}^{\lfloor (1-\frac{1-\delta}{2}) \cdot 2^{N_k} \rfloor} \int_{\theta_d}^{\theta_{d+1}} \left( x_A - \frac{d+0.5}{2^{N_k}} \right)^2 dx_A \\
    H_{N_k, \delta} &= -\log_2 \sqrt{12 \cdot M_{N_k, \delta} }
\end{split}
\end{equation}
Fig. \ref{Fig_ReducedRange} shows the estimated entropy of the same system in Fig. \ref{Fig_SupRes} but at $\delta = 95\%$ sample space. The UN method excels over the HS structure because it is specifically designed to minimize errors at two ends. By sacrificing 5\% of the sample space - a reasonable engineering trade-off, an increase of 8-9 bits effective resolution or 256x-512x enhancement of precision is feasible with the UN structure.

\section{Practical Considerations \& Applications}
\label{Sec_PracApp}
\begin{figure*}[t]
\begin{center}
\includegraphics[width=\textwidth]{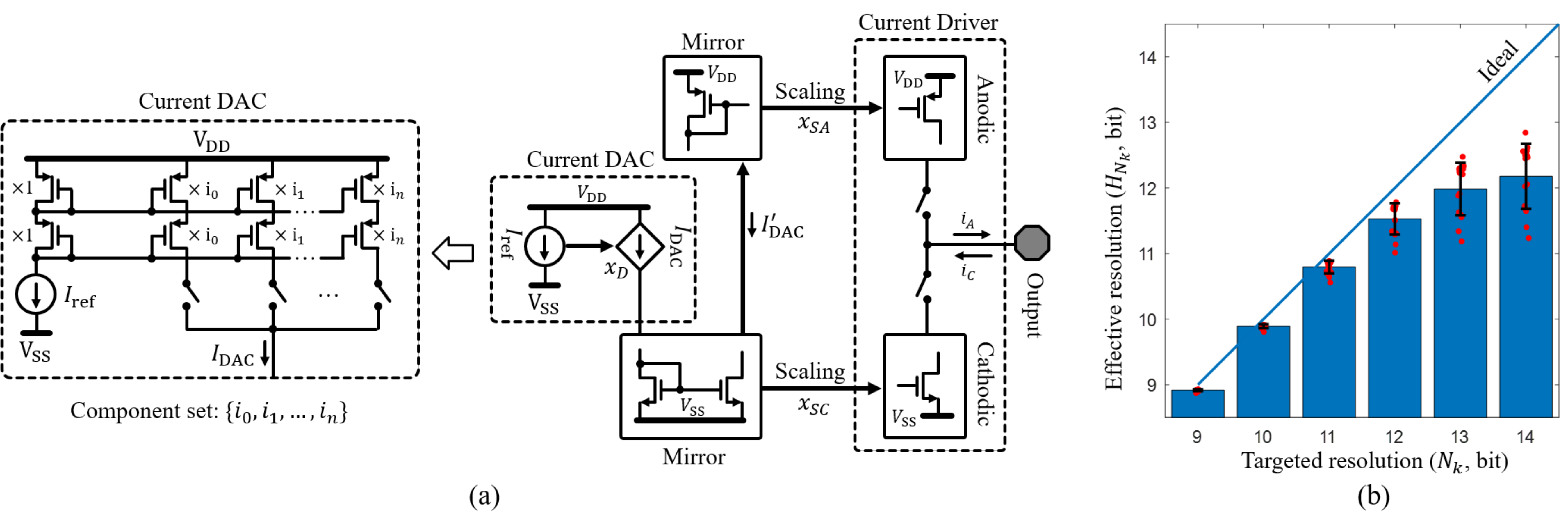}	
\end{center}
\vspace{-0.2in}
\caption{(a) A neurostimulator such as \cite{2016_Nguyen_BioCAS} requires a high-resolution DAC to generate its internal reference current. (b) Monte Carlo simulations at the schematic-level using the transistor's statistical model (both process and variation) show an average of 12-bit effective resolution or a gain of 4-bit extra precision ($N_0 = 8$ bits, $\delta = 95\%$) can be achieved by solely exploiting the natural mismatch of the transistors}
\label{Fig_IDAC}
\end{figure*}

In practice, the greatest challenge for utilizing the proposed SR as well as any RS architecture is to determine the correct configuration of the system among numerous redundant possibilities. In the context of this work, achieving SR at $N_k$ requires solving the following optimization problem:

\textbf{Problem:} \textit{$\forall \theta_i \in \Theta_{N_k}$, find a subset of $C = \{ c_1, c_2, ...c_n \}$ such that it generates a reference $\hat{\theta}_i$ which minimize the error $|\theta_i - \hat{\theta}_i|$}

This is essentially a version of the \textit{0-1 knapsack problem}, which has been shown to NP-hard \cite{2004_Kellerer}. Because of $\Theta_{N_0} \subset \Theta_{N_1} \subset ... \subset \Theta_{N_k}$, achieving SR at any targeted resolution $N_k$ is as hard as the non-SR case of $N_0$ given the actual weights all of the components are known. However, this does not necessarily negate the practicality of the proposed technique. The practical solution to this seemingly unsolvable problem may be specific to each application. 

In \cite{2015_Nguyen}, we have shown an implementation of a high-precision ADC in integrated circuits where a RS structure was utilized. The optimization problem was successfully overcome by employing a heuristic-based approximation algorithm which was derived to compute the near-optimal system configuration on-the-fly given the input signal and the estimated weights of all components. The proposed algorithm was sufficiently simple such that it can be implemented on-chip with an adequate accuracy and a time complexity of merely $O(1)$. This example suggests that approximation would be a viable approach for implementing SR in practice. Realizing SR for ADC would enable enormous boost of performance in various high-precision imaging and instrumentation systems as ADC is one of the core components of many sensors and devices.

SR would also find many uses in DAC devices. For example, an implantable neurostimulator such as \cite{2016_Nguyen_BioCAS} requires a DAC to generate its internal reference current. Higher resolution DAC is always desirable as it gives more precise control of the stimulation current which could imply better modulation of neural circuits. 

Fig. \ref{Fig_IDAC}a shows the functional blocks of the neurostimulator and the schematic of its current DAC where each unit cell is a MOS transistor. Although mostly time-invariant, transistor mismatch is particularly complex because it not only depends on the device's physical size ($W/L$) but also the operating conditions such as biasing voltage, loading current, parasitics, etc. As a proof-of-concept demonstration, we design and simulate a SR DAC in the GlobalFoundries BCDLite 0.18$\mu$m process using 30V transistors with minimum feature size ($W/L = 4.0/0.5\mu m$). The DAC architecture employs the UN grouping and has an intrinsic resolution of $N_0 = 8$ bits. Monte Carlo simulations ($n = 16$) at the schematic-level are performed using the transistor's statistical model (both process and variation) provided by the foundry without any added pseudo-random mismatch. The model should account for the majority of the mismatch except for the parasitic resistance of metal connections in the layout. Fig. \ref{Fig_IDAC}b shows the simulation results where an average of 12-bit effective resolution or a gain of 4-bit extra precision at $\delta = 95\%$ can be achieved by solely exploiting the natural mismatch of the transistors. The results show a concrete example where the proposed SR mechanism can be utilized to greatly enhance the performance of a high-precision device. 

Moreover, unlike the ADC example, the neurostimulator's operations are always governed by an external controller during normal operation. The controller regularly communicates with the neurostimulator to update its parameters and trigger its function when needed. Subsequently, the optimal system setting at every DAC output can be simply determined upfront via foreground calibration and saved on an external memory which is accessed by the controller at any instant. This effectively circumvents the computational-hard problem by diverting it into a memory-hard problem which could be more easily handled in certain circumstances. For instance, assuming a targeted SR of 16-bit is to be achieved with 20 components, storing all the optimal configurations would require $2^{16} \times 20 = 1.3 \cdot 10^6$ bits or 163KB of memory per DAC - a trivial amount for an external flash memory.

\section{Conclusion}
\label{Sec_Conclusion}

This work presents a new interpretation of the RS architecture that allows quantization or de-quantization processes to achieve an effective resolution many folds beyond the limitation that their resource constraints commonly permit. Using Monte Carlo simulations, we show that SR is feasible by elegantly exploiting the statistical property called ``code diffusion'' that is unique to a redundant structure in the presence of random ME. By applying the proposed technique on a 10-bit device, a profound theoretical increase of 8-9 bits effective resolution or 256-512x enhancement of precision at 95\% sample space is demonstrated. We also point out the challenges of utilizing the proposed mechanism in practice but argue that they can be overcome by means of approximation or avoided in certain conditions. We envision the proposed technique would give rise to wide applications in various fields of imaging and data acquisition instrumentation, especially fully-integrated sensors and devices where higher resolution is always desired.

\section*{Acknowledgments}
This work was sponsored by the Defense Advanced Research Projects Agency (DARPA), Biological Technologies Office (BTO), contract No. HR0011-17-2-0060 as well as internal funding from the University of Minnesota.

\end{document}